\documentstyle[aps, prl]{revtex}
\begin{document}
\author{A.S. Alexandrov$^{1}$ and V.V. Kabanov $^{2}$}
\address{$^{1}$ Department of Physics, Loughborough University,\\
Loughborough LE11\\
3TU, U.K.\\
$^{2}$ Bogoliubov Laboratory of Theoretical Physics, JINR, Dubna, Russia;\\
and Josef Stefan Institute SI-1001, Ljubljana, Slovenia.}
\title{Parameter-free expression for superconducting $T_{c}$ in cuprates} 
\maketitle

\begin{abstract}
A parameter-free expression for the superconducting critical temperature of
layered cuprates is derived which allows us to express T$_{c}$ in terms of
experimentally measured parameters. It yields T$_{c}$ values observed in
about 30 lanthanum, yttrium and mercury-based samples for different levels
of doping. This remarkable
agreement with
the experiment as well as  the unusual critical
behaviour and the normal-state gap
  indicate  that many cuprates are close to the Bose-Einstein
 condensation  regime.
\end{abstract}


\twocolumn An ultimate goal of the theory of superconductivity is to provide
an expression for $T_{c}$ as a function of some well-defined parameters
characterizing the material \cite{gin}.  The BCS theory  provides a
`material' aspect in an
estimate of $T_{c}$ since the characteristic
phonon frequency  and the electron-phonon coupling constant
 can be measured \cite{sca}
while the Coulomb pseudopotential  is determined in the first
principle (LDA) band-structure calculations.
There is no general  restrictions on the BCS value of
T$_{c}$ if the dielectric function formalism is properly applied
\cite{max,all}.
   On the other hand, there is a growing evidence for
 a non-Fermi-liquid  normal state and for a  non-BCS
superconducting state of doped cuprates (see, for example \cite{alemot}).
In particular,  the correlation of T$_{c}$ with the carrier
 density and with their effective mass \cite{mih,uem2},  the carrier
specific heat
 near the transition   and its magnetic field dependence \cite{jun},
 the divergent upper critical field \cite{oso,ale3,law} are
 reminicent of a charged Bose liquid \cite{alemot,ale4,aleH}.

 In this letter we derive   a
 parameter-free expression for T$_{c}$ of
 a charged Bose-liquid on a quasi-2D lattice. It
yields  T$_{c}$ values observed in many cuprates for any
 level of doping. The main point of our letter is largely independent of
 the microscopical interpretation of charged bosons, which might be
lattice and/or spin bipolarons \cite{alemot}, or any other preformed pairs.

 In the framework of this rather general model  T$_{c}$
 is calculated from the density sum
 rule as the Bose-Einstein condensation temperature of 2e charged
 bosons on a lattice. Just before the discovery \cite{bed} we
estimated T$_{c}$ as high as $\simeq 100$K   by applying
our expression for the bipolaron effective mass \cite{alekab}.
That was tested by Uemura \cite{uem2,uem} with the
 conclusion that cuprates are neither  BCS nor   BEC superfluids
 but they are in a crossover region from one to another.
 The experimental T$_{c}$ has been  found about 3 or more times
 below the BEC temperature.

 We  now
 calculate the critical temperature of a charged Bose-liquid
 taking carefully into account the microscopic band structure of
 bosons in layered cuprates. It appears that due to a multi-band structure
 bosons have internal quantum
 number that might give a two-fold degeneracy as  derived
  by one of us for a particular case of a `peroxy' bipolaron \cite{ale}. We
arrive at the
parameter-free expression for T$_{c}$, which involves the in-plane, $\lambda
_{ab}$ and out-of-plane, $\lambda _{c}$ magnetic field penetration depths
and the normal state Hall coefficient $R_{H}$ just above the transition. It
describes the experimental data for a few dozen different samples clearly
indicating that many cuprates are in the BEC regime.

 Bound pairs in
cuprates are intersite  pairs \cite{alemot}
because of a strong on-site repulsion.  In the case of the
electron-phonon coupling this is
 confirmed by the numerical simulations of ionic
perovskite lattices \cite{cat}, where two types of pairs were
found, i.e a `peroxy' bipolaron  and an `in-plane' one.  The  energy spectrum
of the 'peroxy bipolaron ( a bound state of the in-plane and apical holes)
is a  doublet due to  two ($x$ and $y$) oxygen
orbitals elongated along the CuO$_{2}$ planes \cite{ale}.
The energy band minima are found at the Brillouin zone
 boundary,  $(\pm \pi,0)$ and  $(0, \pm \pi)$
  owing to the opposite sign of the $pp\sigma$ and $pp\pi$
oxygen  hopping integrals. Near these points an effective mass approximation
is applied with the following result for the $x$ and $y$ bipolaron bands:
\begin{equation}
E_{{\bf k}}^{x,y}={\frac{\hbar ^{2}k_{x,y}^{2}}{{2m_{x}}}}+{\frac{\hbar
^{2}k_{y,x}^{2}}{{2m_{y}}}}+t_{c}[1-cos(k_{z}d)],
\end{equation}
where $d$ is the interplane distance and $t_{c}/2$ is the interplane 
hopping integral. The wave vectors corresponding to the energy minima belong
to the stars with two prongs. Their group has only 1D representations. It
means that the spectrum is degenerate with respect to the number of prongs
of the star. The spectrum Eq.(1) belongs to the star with two prongs, and,
hence it is a two-fold degenerate.

While the X-ray absorption experiments confirm an important role of
apical holes in some cuprates \cite{kar} there are multilayer compounds
with inner layers without apex oxygens. The in-plane p-band
hybridised with  copper might have a
higher energy than other bands and the first to be doped. In those cases
 the in-plane boson ( a bound state of two oxygen  holes in the
$CuO_{2}$ plane) is the ground state with
the energy minimum, which might be found at the $\Gamma $ point of the
Brillouin
  zone. One can apply the method of invariants to derive its spectrum. The
space group of $La_{2-x}Sr_{x}CuO_{4}$ in the tetragonal phase is $%
D_{4h}^{17}$. The wave functions at the $\Gamma $ point transform as 1D $%
A_{1g,u}$, $A_{2g,u}$, $B_{1g,u}$, $B_{2g,u}$ or 2D $E_{g\text{ }}$and $E_{u%
\text{ }}$representations of the point group $D_{4h}$. For the case of $E_{g%
\text{ }}$ (basis functions $xz$ and $yz$) and $E_{u\text{ }}$ (basis
functions $x$ and $y$) the spectrum is expressed as the eigenvalues of the
2x2 matrix. This matrix can be written as a linear combination of Pauli
matrices, $\tau _{i}$. Taking into account that $\tau _{0}$ transforms as $%
A_{1g}$ (basis functions $k_{x}^{2}+k_{y}^{2}$, $k_{z}^{2}$), $\tau _{x}$
transforms as $B_{1g}$ (basis functions $k_{x}^{2}-k_{y}^{2}$), $\tau _{y}$
transforms as $B_{2g}$ (basis functions $k_{x}k_{y}$), $\tau _{z}$
transforms as $A_{2g}$ (basis functions $s_{z}$) we write the matrix of the
Hamiltonian as $H=\left[ A(k_{x}^{2}+k_{y}^{2})+Bk_{z}^{2}\right] \tau
_{0}+C(k_{x}^{2}-k_{y}^{2})\tau _{x}+Dk_{x}k_{y}\tau _{y}$. The eigenvalues
of this Hamiltonian are $E_{{\bf k}%
}^{1,2}=A(k_{x}^{2}+k_{y}^{2})+Bk_{z}^{2})\pm \sqrt{%
C^{2}(k_{x}^{2}-k_{y}^{2})^{2}+D^{2}k_{x}^{2}k_{y}^{2}}.$ Here $A$, $B$, $C$%
, and $D$ are fenomenological constants that parametrize the effective mass
tensor. As a result the in-plane spectrum
is degenerate in $\Gamma $ point as well if it belongs to a 2D
representation.

It should be pointed out that a low temperature phase has lower symmetry
(space group $D_{2h}^{18}$). It means that the  degeneracy of the spectrum
might be removed below the structural phase transition. Nevertheless, if
the level splitting is less or of the order of $T_{c}$ we can
 apply a theory
with a two-fold degenerate spectrum. On the other hand, if the degeneracy
is removed the theoretical T$_{c}$ would be higher by $2^{2/3}$ (see
below).

Substituting the spectrum, Eq.(1) into the density sum rule,
\begin{equation}
\sum_{{\bf k},i=(x,y)}\left[ \exp (E_{{\bf k}}^{i}/k_{B}T_{c})-1\right]
^{-1}=n_{B}
\end{equation}
one readily obtains $T_{c}$ as
\begin{equation}
k_{B}T_{c}=f\left( {\frac{t_{c}}{{k_{B}T_{c}}}}\right) \times {\frac{%
3.31\hbar ^{2}(n_{B}/2)^{2/3}}{{(m_{x}m_{y}m_{c})^{1/3}}}},
\end{equation}
where the coefficient $f\simeq 1$ is shown in Fig.1 as a function of the
anisotropy, $t_{c}/k_{B}T_{c}$, and $m_{c}=\hbar ^{2}/|t_{c}|d^{2}$.This
expression is rather ambiguous so far because  the effective mass
tensor as well as the boson density n$_{B}$ are unknown and doping
 dependent due to the screening of the interaction by free
carriers and their localisation by disorder.
  Fortunately, one  can express the band-structure parameters
through the in-plane,
 $\lambda_{ab}=[m_{x}m_{y}/8\pi
 n_{B}e^{2}(m_{x}+m_{y})]^{1/2}$ and out-of-plane penetration depth,
 $\lambda_{c}=[m_{c}/16\pi
 n_{B}e^{2}]^{1/2}$ (we take $c=1$). The boson density  is
 expressed through the in-plane Hall constant (above the transition) as
 \cite{ale}:
 \begin{equation}
 R_{H}={1\over{2en_{B}}}\times {4m_{x}m_{y}\over{(m_{x}+m_{y})^{2}}},
 \end{equation}
which leads to
\begin{equation}
T_{c}=1.64 f\times \left( {\frac{eR_{H}}{{\lambda _{ab}^{4}\lambda _{c}^{2}}}}%
\right) ^{1/3}
\end{equation}
with T$_{c}$ measured in Kelvin, $eR_{H}$ in cm$^{3}$ and $\lambda $ in cm.
The coefficient $f$ is about unity in a very wide range of the anisotropy $%
t_{c}/k_{B}T_{c}\geq 0.01$, Fig.1. As a result we arrive at a parameter-free
expression, which unambiguously
can tell us how far cuprates are from the BEC regime:
\begin{equation}
T_{c}\simeq T_{c}(3D)=1.64
\left({eR_{H}\over{\lambda_{ab}^{4}\lambda_{c}^{2}}}\right)^{{1/3}}.
\end{equation}
It does not contain the mass tensor explicitly. Hence, any other
dispersion law would lead to about the same result for T$_{c}$ if the spectrum
is two-fold degenerate. As an example, taking for simplicity $C=D=0$ in the
in-plane
bound state dispersion,   one arrives at $T_{c}\simeq 3.31
(n_{B}/2)^{2/3}/(m^{2} m_{c})^{1/3}$, $R_{H}=1/2en_{B}$, and
$\lambda_{ab}=[m_{ab}/16\pi
 n_{B}e^{2}]^{1/2}, \lambda_{c}=[m_{c}/16\pi
 n_{B}e^{2}]^{1/2}$, and with Eq.(6) as well. Here $m_{ab}=\hbar^{2}/2A$
 and $m_{c}=\hbar^{2}/2B$.

 We compare the theoretical expressions, Eq.(5,6) with the experimental
T$_{c}$ of
 more than 30 different cuprates, for which both $\lambda_{ab}$ and
 $\lambda_{c}$ are measured along with
 $R_{H}$ (see  Table 1 and  Fig.2). The Hall coefficient has a strong
 temperature dependence above T$_{c}$. Therefore, we use the experimental Hall
 `constant' $R_{H}\equiv R_{H}(T_{c}+0)$ just above the transition as
 prescribed by Eqs.(5,6).
 In a few cases (mercury compounds), where R$_{H}(T_{c}+0)$ is unknown, we
have taken the
inverse chemical density of carriers (divided by e) as R$_{H}$. Almost for
 all  samples the  theoretical T$_{c}$
coincides with the experimental one  within the experimental error
bar  for the penetration depth (about $\pm 10 \%$).
There are a few Zn doped YBCO samples, Fig.2,  with the experimental critical
temperature higher than the theoretical one. We believe that
 the degeneracy of the boson spectrum is removed by the
random potential of Zn, so for these samples
the theoretical T$_{c}$ is actually higher
than Eq.(6) suggests. Multiplying the theoretical T$_{c}$ in Table 1 by
$2^{2/3}$ for three Zn-doped samples we obtain T$_{c}$=73 K, 57 K, 41 K in
 good agreement with the experimental values T$_{c}$=68 K, 55 K
and 46 K, respectively.

One can argue that cuprates belong to a 2D `XY' universality class with
the Kosterlitz-Thouless critical temperature T$_{KT}$ of preformed
bosons \cite{ald,poc} or Cooper pairs \cite{eme}
due to a large anisotropy. Would it be the case
 one could hardly discriminate
Cooper pairs with respect to local pairs in cuprates by the use of their T$_{c}$
values. The Kosterlitz-Thouless temperature is
expressed trough the in-plane penetration depth alone  as \cite{eme}
\begin{equation}
k_{B}T_{KT}\simeq {0.9 d \hbar^{2}\over{16\pi
e^{2}\lambda_{ab}^{2}}}.
\end{equation}
It appears significantly higher than the
experimental value in most cases (see Table 1).
 What is more crucial, however, is the fact that
cuprates have   the specific heat \cite{jun}
 of a 3D charged Bose gas \cite{ale4}.

The boson-boson interaction might be rather strong leading to
self-energy effects  and to some renormalisation of the effective mass tensor.
It is important, that Eq.(6) does not contain the mass and, hence, is not
 affected by the interaction. Nevertheless, it is interesting to
 evaluate the effective mass tensor in terms of the penetration depth and
 the Hall constant. The in-plane and out-of-plane boson masses are presented in
 Table 2 for a few samples of La$_{2-x}$Sr$_{x}$Cu$_{4}$ and
 YBa$_{2}$Cu$_{3}$O$_{7-x}$. The in-plane boson dispersion with $C=D=0$
 has been applied.
  Theoretical estimates
 of the in-plane mass ($\simeq 10m_{e}$ i.e.about 5m$_{e}$ per hole
 \cite{ale,puc,ald2,cat,alecom})
 fit well our empirical values, Table 2. There is an intersting opposite
 tendency in the doping dependence of the effective mass  of
 La$_{2-x}$Sr$_{x}$CuO$_{4}$ and
 YBa$_{2}$Cu$_{3}$O$_{7-x}$. While  the mass
 increases with doping in La$_{2-x}$Sr$_{x}$CuO$_{4}$, it slightly decreases in
 YBa$_{2}$Cu$_{3}$O$_{7-x}$. We believe that it is a result of an interplay
 between an interaction responsible for the mass enhancement and disorder.
We notice, however,  that the absolute
 value of the effective mass in terms of the free electron mass, Table 2,
does not
describe  the actual band mass renormalisation because the unrenormalised
(bare) band mass remains unknown.

   Many  thermodynamic, magnetic and kinetic properties of
cuprates were understood with charged bosons on a lattice
\cite{alemot,zha}.  We admit, however, that one experimental fact might be
sufficient to destroy any theory.
  In particular, the single-particle
spectral function seen by ARPES \cite{she} was interpreted by several
authors as a
Fermi liquid feature of the normal state incompatible with bipolarons.
Most (but not all) of these measurements produced a large Fermi
surface. This should evolve with doping as $(1-x)$ in a clear
contradiction with low frequency kinetics and thermodynamics, which show
an evolution proportional to $x$ ($x$ is
the number of holes introduced by doping). Recently it has been
established, however, that there is a normal state gap in ARPES and SIN
tunnelling, existing well above
T$_{c}$ irrespective to the doping level \cite{she,bia,ren}. 'The
`Fermi surface' showed missing segments just near the points\cite {bia}
where we
expect the Bose-Einstein condensation \cite{ale2}. A plausible explanation
is that
there are two liquids in cuprates, the normal Fermi liquid and the
charged Bose-liquid (this mixture has been theoretically discussed a long time
ago \cite{alexme}). However, it is difficult to see how this  scenario
could explain the doping dependence of  $dc$ and $ac $
conductivity as well as of the magnetic susceptibility and carrier specific
heat. On the other hand, the single-particle spectral function of a
pure `bosonic'  system has been recently  derived by one of us \cite{ale2}.
It describes the  spectral features of tunnelling
and photoemission in cuprates. Any single-particle
spectral weight at the chemical potential appears in our model  due to
states localised by disorder
inside the normal state gap  \cite{ale2}. The model is thus
compatible with the doping evolution of thermodynamic and kinetic properties.

 In conclusion, we have shown that  the experimental critical
 temperate of superconducting cuprates is not very different from the
 Bose-Einstein condensation temperature of  two-fold degenerate
 charged bosons on a lattice. Our empirical expression for
 T$_{c}$ describes the experimental data remarkably well with no
parameters to fit.
 This  possibility originates in two different energy scales in
 cuprates: a strong attractive interaction and a small bandwidth. Their
 difference allows us to `integrate out' the  interaction
 and express T$_{c}$ via static response functions.

We would like to thank J. Hofer, C. Panagopoulos and V. Pomjakushin for
providing us
with the carrier densities and the penetration depths in several
cuprates and E. Dagotto, L.P. Gor'kov, Guo-meng Zhao,  E.G. Maximov, D.
Mihailovi'c,  S. von Molnar, and
J.R. Schrieffer for enlightening discussions.
One of us (VVK) acknowledges support of
this work by RFBR Grant 97-02-16705 and by the Ministry of Science and
Technology of Slovenia.

\newpage \centerline{{\bf Figure Captures}}

Fig.1. Correction coefficient to the 3D Bose-Einstein condensation
temperature as a function of the anisotropy.

Fig.2. Theoretical critical temperature compared with the experiment ( the
theory is exact for samples on the straight line) for $LaSrCuO$ compounds
(squares) for $Zn$ substituted YBa$_{2}$(Cu$_{1-x}$Zn$_{x}$)$_{3}$O$_{7}$
(circles)
for YBa$_{2}$Cu$_{3}$O$_{7-\delta }$ (triangles) and for HgBa$_{2}$CuO$%
_{4+\delta }$ (diamonds). Experimental data for the London penetration depth
are taken from T. Xiang $et$ $al$, Int. J. Mod. Phys. B ${\bf 12}$, 1007
(1998) and B. Janossy $et$ $al$, Physica C, ${\bf 181}$, 51 (1991) in YBa$%
_{2}$Cu$_{3}$O$_{7-\delta }$ and YBa$_{2}$(Cu$_{1-x}$Zn$_{x}$)$_{3}$O$_{7}$;
from
V.G. Grebennik $et$ $al$, `Hyperfine Interactions', ${\bf 61}$, 1093 (1990)
and C. Panagopoulos (private communication) in underdoped and overdoped La$%
_{2-x}$Sr$_{x}$CuO$_{4}$, respectively, and from J. Hofer $et$ $al$, Physica
C, ${\bf 297}$, 103 (1998) in HgBa$_{2}$CuO$_{4+\delta }$. The Hall
coefficient above T$_{c}$ is taken from A. Carrington $et$ $al$, Phys. Rev.
B ${\bf 48}$, 13051 (1993) and J. R. Cooper (private communication) (YBa$%
_{2} $Cu$_{3}$O$_{7-\delta }$ and YBa$_{2}$(Cu$_{1-x}$Zn$_{x}$)$_{3}$O$_{7}$)
and
from H.Y. Hwang $et$ $al$, Phys. Rev. Lett. ${\bf 72}$, 2636 (1994) (La$%
_{2-x}$Sr$_{x}$CuO$_{4}$).

\begin{table}[tbp]
\caption{Experimental data on $T_{c}$(K), $ab$ and $c$ penetration depth($nm$%
), Hall coefficient ($10^{-3}(cm^{3}/C)$), and calculated values of $T_{c}$
applying Eqs.(6,5,7) respectively (K)for $La_{2-x}Sr_{x}CuO_{4}$ ($La$), $%
YBaCuO(x\%Zn)$ ($Zn$), $YBa_{2}Cu_{3}O_{7-x}$ ($Y$) and $HgBa_{2}CuO_{4+x}$ (%
$Hg$) compounds}
\begin{tabular}[t]{llllllll}
Compound & $T_{c}$ & $\lambda_{ab}$ & $\lambda_{c}$ & $R_{H}$, & $T_{c}$ & $%
T_{c}$ & $T_{KT}$ \\ \hline
$La$(0.2) & 36.2 & 200 & 2540 & 0.8 & 38 & 41 & 93 \\
$La$(0.22) & 27.5 & 198 & 2620 & 0.62 & 35 & 36 & 95 \\
$La$(0.24) & 20.0 & 205 & 2590 & 0.55 & 32 & 32 & 88 \\
$La$(0.15) & 37.0 & 240 & 3220 & 1.7 & 33 & 39 & 65 \\
$La$(0.1) & 30.0 & 320 & 4160 & 4.0 & 25 & 31 & 36 \\
$La$(0.25) & 24.0 & 280 & 3640 & 0.52 & 17 & 19 & 47 \\
$Zn$(0) & 92.5 & 140 & 1260 & 1.2 & 111 & 114 & 172 \\
$Zn$(2) & 68.2 & 260 & 1420 & 1.2 & 45 & 46 & 50 \\
$Zn$(3) & 55.0 & 300 & 1550 & 1.2 & 35 & 36 & 38 \\
$Zn$(5) & 46.4 & 370 & 1640 & 1.2 & 26 & 26 & 30 \\
$Y$(0.3) & 66.0 & 210 & 4530 & 1.75 & 31 & 51 & 77 \\
$Y$(0.43) & 56.0 & 290 & 7170 & 1.45 & 14 & 28 & 40 \\
$Y$(0.08) & 91.5 & 186 & 1240 & 1.7 & 87 & 88 & 98 \\
$Y$(0.12) & 87.9 & 186 & 1565 & 1.8 & 75 & 82 & 97 \\
$Y$(0.16) & 83.7 & 177 & 1557 & 1.9 & 83 & 89 & 108 \\
$Y$(0.21) & 73.4 & 216 & 2559 & 2.1 & 47 & 59 & 73 \\
$Y$(0.23) & 67.9 & 215 & 2630 & 2.3 & 46 & 58 & 73 \\
$Y$(0.26) & 63.8 & 202 & 2740 & 2.0 & 48 & 60 & 83 \\
$Y$(0.3) & 60.0 & 210 & 2880 & 1.75 & 43 & 54 & 77 \\
$Y$(0.35) & 58.0 & 204 & 3890 & 1.6 & 35 & 50 & 82 \\
$Y$(0.4) & 56.0 & 229 & 4320 & 1.5 & 28 & 42 & 65 \\
$Hg$(0.049) & 70.0 & 216 & 16200 & 9.2 & 23 & 60 & 115 \\
$Hg$(0.055) & 78.2 & 161 & 10300 & 8.2 & 43 & 92 & 206 \\
$Hg$(0.055) & 78.5 & 200 & 12600 & 8.2 & 28 & 69 & 134 \\
$Hg$(0.066) & 88.5 & 153 & 7040 & 6.85 & 56 & 105 & 229 \\
$Hg$(0.096) & 95.6 & 145 & 3920 & 4.7 & 79 & 120 & 254 \\
$Hg$(0.097) & 95.3 & 165 & 4390 & 4.66 & 61 & 99 & 197 \\
$Hg$(0.1) & 94.1 & 158 & 4220 & 4.5 & 66 & 105 & 216 \\
$Hg$(0.101) & 93.4 & 156 & 3980 & 4.48 & 70 & 107 & 220 \\
$Hg$(0.101) & 92.5 & 139 & 3480 & 4.4 & 88 & 127 & 277 \\
$Hg$(0.105) & 90.9 & 156 & 3920 & 4.3 & 69 & 106 & 220 \\
$Hg$(0.108) & 89.1 & 177 & 3980 & 4.2 & 58 & 90 & 171 \\ \hline
\end{tabular}
\end{table}

\begin{table}[tbp]
\caption{Mass enhancement with respect to the free electron
mass}
\begin{tabular}[t]{lll}
Compound & $m_{ab}$ & $m_{c}$ \\ \hline
$La$(0.2) & 22.1 & 3558 \\
$La$(0.15) & 15.0 & 2698 \\
$La$(0.1) & 11.3 & 1909 \\
$Y$(0.0) & 7.2 & 584 \\
$Y$(0.12) & 8.3 & 600 \\
$Y$(0.3) & 10.6 & 1994 \\
\hline
\end{tabular}
\end{table}

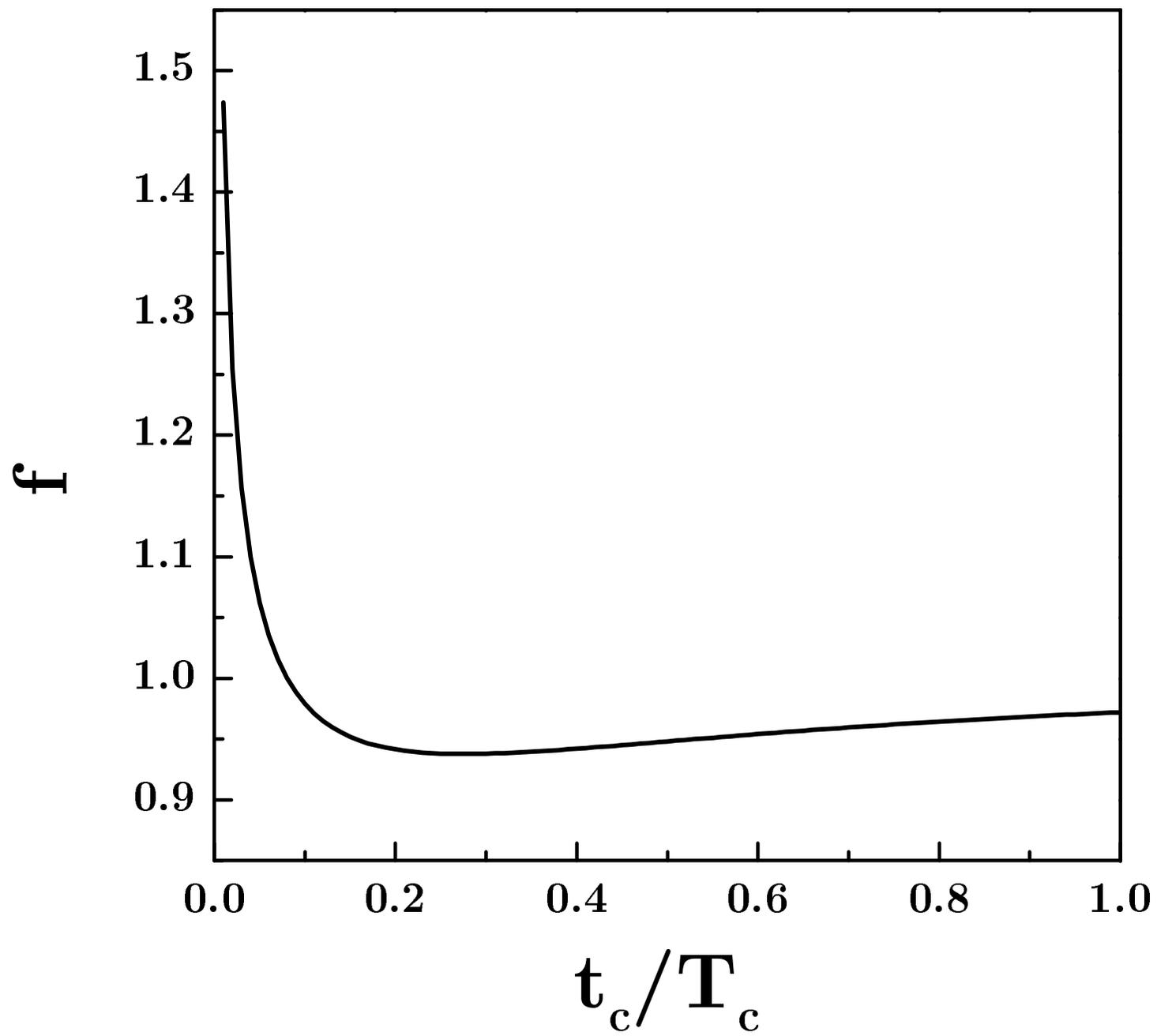
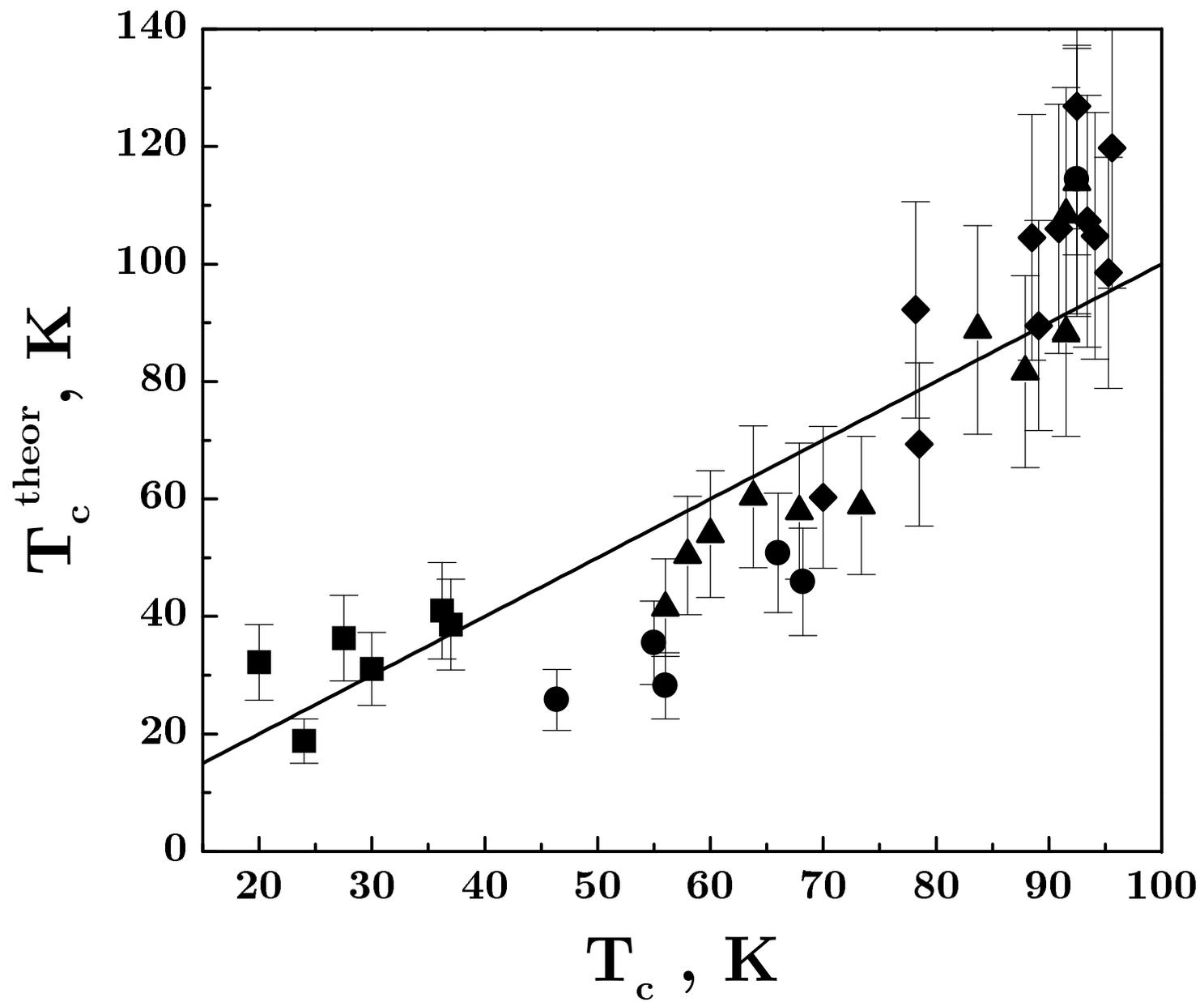
\end{document}